\documentclass{article}
\usepackage{graphicx} 

\usepackage{fullpage,amsmath,amsthm,amssymb,enumerate,enumitem,mathtools,float,accents,url,enumerate, tikz, subcaption,comment}

\usepackage[round]{natbib}
\usepackage[colorlinks = true, citecolor = blue,
linkcolor = blue]{hyperref}
\usepackage{cleveref}
\usepackage{parskip}
\usepackage{setspace}

\usepackage{authblk}

\DeclareMathOperator*{\argmin}{argmin}

\title{Two Approaches to Direct Estimation of Riesz Representers}
\author[]{David Bruns-Smith}
\affil[]{Stanford University}
\date{}

\begin{document}

\maketitle

\begin{abstract}
    The Riesz representer is a central object in semiparametric statistics and debiased/doubly-robust estimation. Two literatures in econometrics have highlighted the role for directly estimating Riesz representers: the automatic debiased machine learning literature \citep[as in][]{chernozhukov2022automatic}, and an independent literature on sieve methods for conditional moment models \citep[as in][]{chen2014sieve}. These two literatures solve distinct optimization problems that in the population both have the Riesz representer as their solution. We show that with unregularized or ridge-regularized linear, sieve, or RKHS models, the two resulting estimators are numerically equivalent. However, for other regularization schemes such as the Lasso, or more general machine learning function classes including neural networks, the estimators are not necessarily equivalent. In the latter case, the \cite{chen2014sieve} formulation yields a novel constrained optimization problem for directly estimating Riesz representers with machine learning. Drawing on results from \cite{birrell2022optimizing}, we conjecture that this approach may offer statistical advantages at the cost of greater computational complexity.
\end{abstract}

\section{Introduction}







The Riesz representer arises as a key object in semiparametric statistics and debiased/doubly-robust estimation. For causal inference researchers, it is easiest to think of the Riesz representer as a generalization of the inverse probability weights (IPW) --- see \cite{williams2025riesz} for a review. There has been substantial recent interest in directly estimating Riesz representers, as opposed to e.g. estimating the propensity score and then inverting \citep{chernozhukov2022automatic, lee2025rieszboost}. Direct estimation has statistical advantages, applies ``automatically'' to a large class of estimands, and is amenable to estimation with machine learning. In computer science, an identical approach was proposed in \cite{kanamori2008efficient}.\footnote{This connection is not necessarily obvious. In \cite{kanamori2008efficient}, the object of interest was the density ratio between a source and target distribution. Every such density ratio is a Riesz representer for the mean under the target distribution. Conversely, every Riesz representer can be written as a Radon-Nikodym derivative where the target is a finite signed measure.} These approaches have deep roots in the survey calibration literature \citep{deville1992calibration} and have a primal/dual relationship with ``balancing'' estimators \citep{graham2012inverse, hainmueller2012entropy, ben2021balancing}. See \cite{bruns2025augmented} for further discussion.

A largely separate literature in econometrics on sieve methods in conditional moment models has also explicitly emphasized the role of the Riesz representer in the efficient asymptotic variance going back to the 90's \citep{shen1997methods, chen1998sieve, ai2003efficient, ai2007estimation}. This literature also developed direct estimators for the Riesz representer \citep{chen2014sieve, chen2015sieve} that solve a slightly different optimization problem than the one used in \cite{chernozhukov2022automatic}. A closed-form for the unregularized case (with the optimization problem left implicit) was applied earlier to density ratio estimation in \cite{chen2005measurement}.

The purpose of this note is to document some simple new results comparing the approaches of these two literatures:
\begin{enumerate}
    \item With an unregularized or ridge-regularized linear, sieve, or RKHS model for the Riesz representer, the estimators in \cite{chen2014sieve} and \cite{chernozhukov2022automatic} are numerically-equivalent. 
    \item For other choices of regularization such as the Lasso, or more general machine learning models including neural networks, the estimators are not necessarily numerically-equivalent.
    \item Finally, we conjecture based on results from \cite{birrell2022optimizing} that implementing the approach in \cite{chen2014sieve} with machine learning could offer statistical advantages at the cost of additional computational complexity. 
\end{enumerate}

\section{Two Optimization Problems for the Riesz Representer}\label{sec:two-problems}

We first describe the key difference in approach between \cite{chernozhukov2022automatic} and \cite{chen2014sieve} in a very generic setup. Both approaches use the fact that the Riesz representer can be characterized as the solution to an optimization problem. They differ in the choice of optimization problem.

Let $\mathcal{H}$ be a Hilbert space. Consider a continuous linear functional $L(h)$ on $h \in \mathcal{H}$. This has a Riesz representer $\alpha_0$ such that $L(h) = \langle h, \alpha_0 \rangle, \forall h \in \mathcal{H}$. The Riesz representer can be written as the solution to the following two optimization problems.

The version used in \cite{sugiyama2010conditional} and \cite{chernozhukov2022automatic} is:

\[ \alpha_0 = \argmin_{\alpha \in \mathcal{H}} \{ \Vert \alpha \Vert^2 - 2 L(\alpha) \}, \]

with the minimum value equal to $- \Vert \alpha_0 \Vert^2$. Chernozhukov and co-authors call this the ``Riesz loss''.

\cite{chen2014sieve, chen2015sieve} solve the optimization problem:

\[ \Vert \alpha_0 \Vert^2 = \max_{\alpha \neq 0} \frac{L(\alpha)^2}{\Vert \alpha \Vert^2}, \]

where $\alpha_0$ is the solution with the appropriate norm. 

\section{Solving the Sample Problem with Sieves}

\subsection{Setup}

The above setup is highly abstract. Now consider a statistical setting with random variables $X \in \mathcal{X}$ and $Y \in \mathbb{R}$. Let $\mathcal{H} = L_2(X)$ and consider an estimand $L(h_0)$, where $h_0(x) \coloneqq \mathbb{E}[Y|X=x]$ and where $L(h) \coloneqq \mathbb{E}[m(h; X)]$ for some $m$ making $L$ a continuous linear functional. One concrete example of such an estimand would be the average treatment effect where $X = (T,W)$ for treatment $T$ and covariates $W$, and
\[ L(h) = \mathbb{E}[ h(1,W) - h(0,W)].\]
In this case, the Riesz representer of $L$ would involve the typical inverse probability weights:
\[ \alpha_0(X) = \frac{T}{\pi(W)} - \frac{1-T}{1-\pi(W)}. \]
In this note, we let $h_0$ be the conditional mean for simplicity, but in \cite{chen2014sieve} this could be the solution to a conditional moment equality. 

We now consider solving the optimization problems from the previous section, using $n$ iid observations of $X$. We will solve the optimization problems over linear functions $h(x) = \theta^T \phi(x)$ for some transformation $\phi : \mathcal{X} \rightarrow \mathbb{R}^d.$ We write $\Phi \in \mathbb{R}^{n \times d}$ for the matrix with rows $\phi(x)$ for each observation.

For such linear functions we have the convenient form:

\[ L(h) = \theta^T \mathbb{E}[ m(\phi ; X) ]. \]

Write $\hat{L}(\Phi) \coloneqq \hat{\mathbb{E}}[ m(\phi ; X) ] \in \mathbb{R}^d$. 

\subsection{Equivalence Without Regularization}

In this setting, the sample version of the two optimization problems in \Cref{sec:two-problems} are numerically-identical.

The version from \cite{sugiyama2010conditional, chernozhukov2022automatic} is:
\begin{align}
    \min_{\theta} \theta^T\Phi^T \Phi \theta - 2 \theta^T \hat{L}(\Phi),\label{eq:sieve-riesz}
\end{align}
which, from the first order conditions, has the closed form solution:
\[ \theta^* = (\Phi^T \Phi)^{-1} \hat{L}(\Phi). \]
As discussed in \cite{bruns2025augmented}, if $\Phi^T \Phi$ is not invertible, then replacing the inverse with the \emph{pseudoinverse} will yield the minimum-norm solution.

The version from \cite{chen2014sieve, chen2015sieve} is now:
\begin{align}
    \max_{\theta \neq 0} \frac{\theta^T  \hat{L}(\Phi)  \hat{L}(\Phi)^T \theta}{\theta^T \Phi^T \Phi \theta}.\label{eq:sieve-rayleigh}
\end{align}
We can rewrite this as the constrained optimization problem:
\[ \max_{\theta : \theta^T \Phi^T \Phi \theta = 1 } \theta^T  \hat{L}(\Phi) \hat{L}(\Phi)^T \theta. \]
Using the Lagrangian, plus the fact that $\hat{L}(\Phi)  \hat{L}(\Phi)^T$ is rank-1, we get that the solutions are: 
\[ \theta_* = (\Phi^T \Phi)^{-1}  \hat{L}(\Phi) \] 
and any positive scalar multiple thereof. A quick check shows that $(\Phi^T \Phi)^{-1}  \hat{L}(\Phi)$ is the solution whose norm is equal to the maximum value. Thus, at least with unregularized linear models, the sample optimization problems have numerically-identical solutions. In fact, this same solution is used in equation (10) of \cite{chen2005measurement}, an application to direct density-ratio estimation that predates \cite{kanamori2008efficient}.

\subsection{Adding Regularization}

Adding regularization to the Riesz loss in \eqref{eq:sieve-riesz} is straightforward: we add a penalty on $\theta$ in a chosen norm, such as the $\ell_1$- or $\ell_2$-norm. Similarly, we could directly add a regularization penalty on $\theta$ to Problem \eqref{eq:sieve-rayleigh}. In the special case of $\ell_2$-norm regularization, the two optimization problems are still equivalent and have solution:
\[ \theta^* = (\Phi^T \Phi + \lambda I)^{-1} \hat{L}(\Phi). \]
By contrast, with $\ell_1$-norm penalties, these optimization problems are no longer equivalent. 

Taking inspiration from the optimization literature, there are other ways to regularize \eqref{eq:sieve-rayleigh}. In particular, the objective in \eqref{eq:sieve-rayleigh} is an example of a \emph{generalized Rayleigh quotient}. An existing literature in machine learning and optimization considers regularized solutions to problems like \eqref{eq:sieve-rayleigh}. See for example \cite{mahoney2010implementing}. This literature reformulates \eqref{eq:sieve-rayleigh} as a semi-definite program (SDP), which introduces natural forms of regularization beyond just the norm of $\theta$. In future work it may be interesting to explore this connection further in the context of estimating Riesz representers. 

\section{Solving the Sample Problem With Machine Learning}

Beyond linear bases, we could also solve the sample optimization problems using machine learning. Let $\mathcal{F}$ represent a machine learning function class such as trees or neural networks. Then we can minimize the Riesz loss over this function class as in \cite{chernozhukov2022riesznet}:

\begin{align}
    \min_{\alpha \in \mathcal{F}} \{ \hat{\mathbb{E}}[ \alpha(X)^2 - 2 m(\alpha ; X) ] \}.\label{eq:ml-riesz}
\end{align}

For example, \cite{lee2025rieszboost} implements this approach with gradient-boosted trees.

We could also use machine learning to solve an analog of Problem \eqref{eq:sieve-rayleigh} from \cite{chen2014sieve}:

\begin{align}
    \max_{\alpha \in \mathcal{F} : \hat{\mathbb{E}}[\alpha(X)^2] = 1} \hat{\mathbb{E}}[ m(\alpha ;X) ]^2.\label{eq:ml-rayliegh}
\end{align}

For an arbitrary machine learning function class $\mathcal{F}$, the solution to these two optimization problems will generally differ. Therefore, \eqref{eq:ml-rayliegh} provides a novel way to directly estimate Riesz representers with machine learning. While we defer a statistical analysis of \eqref{eq:ml-rayliegh} to future work, we make two speculative comments in this section:
\begin{enumerate}
    \item Problem \eqref{eq:ml-rayliegh} will be computationally more complex to solve than minimizing the Riesz loss. 
    \item However, Problem \eqref{eq:ml-rayliegh} may offer some statistical efficiency gains. 
\end{enumerate}

Computationally, the Riesz loss minimization problem \eqref{eq:ml-riesz} is quadratic and unconstrained, making it especially amenable to solution with machine learning methods. By contrast, \eqref{eq:ml-rayliegh} is a constrained optimization problem. It could be implemented with projected gradient descent, or similarly, we could normalize the function before computing the objective and backpropagate gradients through the normalization step. While these ideas are conceptually straightforward to implement with neural networks, training in practice may be more challenging. 

In exchange for solving a constrained (as opposed to unconstrained) optimization problem, we conjecture that there may be gains in statistical estimation. In particular, we highlight a surprising connection between the optimization problem \eqref{eq:ml-rayliegh} inspired by \cite{chen2014sieve} and the GAN training literature in machine learning. Consider the special case where we are estimating a missing mean under covariate shift $\mathbb{E}_Q[Y]$ given data from $P$. In this setting, the Riesz representer is the density ratio $dQ/dP$. The paper \cite{birrell2022optimizing} considers using ``variational representations'' of $f$-divergences to estimate $dQ/dP$. They derive two variational representations for the $\chi^2$-divergence which happen to be exactly equal to the two optimization problems described in \Cref{sec:two-problems}.

This is a potentially interesting connection because \cite{birrell2022optimizing} claims that at least for density ratios, the problem in \cite{chen2014sieve, chen2015sieve} is ``tighter'' (in a particular formal sense) than the one in \cite{chernozhukov2022automatic}, leading to improved statistical estimation. So while the Rayleigh-quotient-type optimization problem may be more challenging to solve computationally with machine learning, this may nonetheless be a promising direction for future work. 

\clearpage
\bibliographystyle{abbrvnat}
\bibliography{var}

\end{document}